# Photovoltaic effect in few-layer black phosphorus PN junctions defined by local electrostatic gating

Michele Buscema,[†] Dirk J. Groenendijk[†], Gary A. Steele, Herre S.J. van der Zant and Andres Castellanos-Gomez*

Kavli Institute of Nanoscience, Delft University of Technology, Lorentzweg 1, 2628 CJ Delft

(The Netherlands).

A.CastellanosGomez@tudelft.nl

The photovoltaic effect is one of the fundamental light-matter interactions in light energy harvesting. In conventional photovoltaic solar cells, the photogenerated charge carriers are extracted by the built-in electric field of a PN junction, typically defined by ionic dopants in a semiconductor. In atomically thin semiconductors, the doping level can be controlled by the field-effect without need of implanting dopants in the lattice, which makes 2D semiconductors prospective materials to implement electrically tunable PN junctions. However, most 2D semiconducting materials do not show ambipolar P-type and N-type field-effect transport, necessary to realize PN junctions. Few-layer black phosphorus is a recently isolated 2D semiconductor that presents a direct bandgap, high mobility, current on/off ratio and ambipolar operation. Here, we fabricate few-layer black phosphorus (b-P) field-effect transistors with split bottom gates and crystalline hexagonal boron nitride (h-BN) dielectric. We demonstrate electrostatic control of the local charge carrier type and density above each gate in the device, tuning its electrical behaviour from metallic to rectifying. Illuminating a gate-defined PN junction, we observe zero-bias photocurrents and significant open-circuit voltage, which we attribute to the separation of electron-hole pairs driven by the internal electric field at the junction region. Due to the small bandgap of the material, we observe photocurrents and photovoltages for illumination wavelengths up to 940 nm, attractive for energy harvesting applications in the near-infrared region.

In the past few years the research effort on two-dimensional (2D) materials has rapidly increased, driven by the great variety of layered materials and their exceptional properties when reduced to few atomic layers.[1-3] Due to their remarkable mechanical,[4,5] electrical[6,7] and optical properties[8-10] they have proven to be promising candidates for many applications such as mass sensing,[11] transistor operation,[7] optical communication[12] and photodetection.[13-16] The field of electronics now has access to a large library of 2D materials, ranging from elemental gapless semiconductors (graphene), semiconducting and superconducting chalcogenides (e.g. transition metal dichalcogenides, TMDCs) to large bandgap insulators (e.g. hexagonal boron nitride, h-BN). The combined use of their exceptional properties has recently been made possible due to advancements in transfer techniques.[6,17,18] Different 2D materials can now be used as components of a single device such as a field-effect transistor, where a gate dielectric and a semiconducting channel material are needed. Due to its atomically flat surface and its disorder-free interface with other 2D materials, h-BN is well-suited to be used as a gate dielectric material.[6,19,20] Few-layer black phosphorus (b-P), a recently isolated 2D material, has shown promising characteristics for its use as channel material, such as high mobility, good current ON/OFF ratios and ambipolarity.[21-28] Especially the ambipolarity and the direct bandgap (optimum for visible and near-infrared applications) make b-P an interesting candidate for electrostatically tuneable optoelectronic devices.

Here, we transfer h-BN and few-layer black phosphorus onto a pair of local gates to define PN junctions. The electrostatic control of the charge carrier type in different regions of a few-layer b-P flake allows us to tune the electronic behaviour from metallic to diode-like. Under illumination, the PN junctions show photovoltaic effect, demonstrating that a large internal electric field can be obtained by local gating. Beyond fabricating a locally-gated PN junction based on a van der Waals heterostructure of 2D materials, our results set the ground for further research towards photodetection and energy harvesting based on few-layer black phosphorus devices.

To efficiently control the doping level in different regions of the b-P flake, we fabricate a pair of local gates (Ti/AuPd, 5 nm/25 nm) separated by a 300 nm gap on a $SiO_2$/Si substrate by standard e-beam lithography and metal deposition techniques (Figure 1a). The schematics below each panel of Figure 1 show a cross section of the device after each fabrication step. On top of the local gates, we first transfer a thin (~ 20 nm) h-BN flake for use as gate dielectric (Figure 1b). The surface of the h-BN appears to be free of wrinkles and folds from optical inspection. Then, we transfer a few-layer (~ 6-7 nm) b-P flake on top of h-BN such that it is well centred with respect to the local gates (Figure 1c). Both transfer steps are performed *via* a recently developed deterministic dry-transfer method.[18] The b-P flake is then contacted with two leads (Ti/Au 2 nm/ 60 nm), fabricated by e-beam lithography, metal deposition and lift-off (Figure 1d).

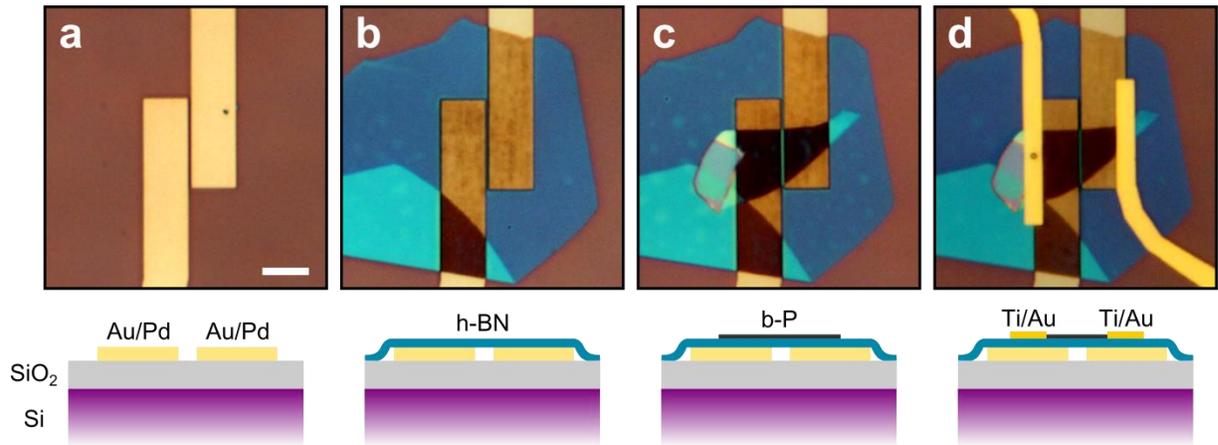

**Figure 1** (**a**) Local gates (Au/Pd) patterned on the surface of a SiO$_2$/Si substrate. The gap between the gates is 300 nm wide. (**b**) After the transfer of a thin h-BN flake on top of the split gates. (**c**) After the transfer of a thin b-P flake on top of the h-BN and the split gate region. (**d**) The final device after fabrication of the contacts (Ti/Au). The schematics below each panel show a cross section of the device after each fabrication step. The scale bar is 5 µm for all panels.

Figure 2a shows the electrical transfer characteristics of a locally gated b-P flake; the source-drain current ($I_{ds}$) is recorded at constant source-drain bias ($V_{ds}$) while the two local gates act as a single back-gate ($V_{lg} = V_{rg} = V_g$). The inset shows the data on a linear scale. The source-drain current is strongly modulated as $V_g$ is swept from negative to positive voltages: $I_{ds}$ varies from a few µA at $V_g$ = -10 V to a minimum of ~ 100 pA at $V_g$ = 5 V and increases to tens of nA for larger $V_g$ values. The non-monotonic transfer curve is a clear signature of ambipolar transport, indicating that both hole and electron doping can be achieved, as expected for few-layer b-P.[26,29] From the figure, we derive that the b-P transistor on h-BN is naturally P-doped with a field-effect mobility in the order of 25 cm$^2$/Vs and current modulation of more than a factor 10$^4$ for holes (0.12 cm$^2$/Vs and 10$^2$ for electrons). We observe similar behaviour in three fabricated locally gated b-P devices. While the results shown in the main text correspond to the same device, the reader can find the characterization and datasets for the other devices in the Supplementary Material. Interestingly, the current values in the hole- and electron-doped regimes at the same carrier concentration differ less than one order of magnitude for all devices. This stronger ambipolar behaviour, in comparison to that observed in back-gated devices on $SiO_2$, is possibly due to the reduced charge transfer provided by the h-BN flake dielectric, which is characterized by a very low density of charged impurities. [21-26,29]

Figure 2b and Figure 2c show the conductance of the device in P- and N-type operation respectively. The measured $I_{ds}$-$V_{ds}$ curves are linear for both operation modes over a wide range of gate voltages, indicating that the Schottky barriers do not dominate the transport and that the Ti/Au leads make good ohmic contact to the b-P flake. The current in the hole-doped regime is higher than in the electron-doped regime, indicating that the Fermi level of the contacts is closer to the valence band than the conduction band of b-P, similar to previous devices contacted with the same metals.[26]

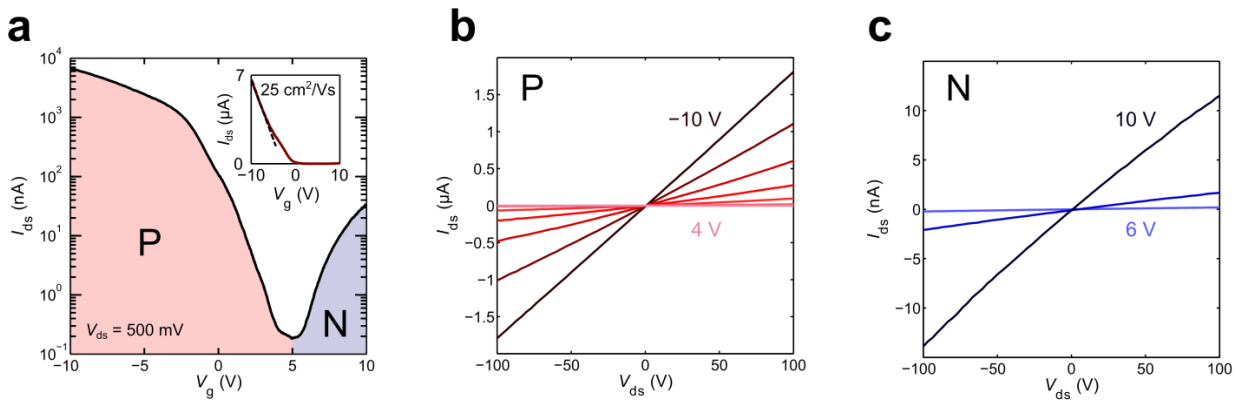

**Figure 2** (**a**) Transfer characteristic at $V_{ds}$ = 500 mV. The inset shows the data in linear scale. (**b**) Output characteristics for $V_g$ from -10 to 4 V in steps of 2 V. (**c**) Output characteristics for $V_g$ from 6 V to 10V in steps of 2 V.

The multiple gates in our device allow us not only to make an ambipolar transistor, such as in Figure 2a, but also to locally induce hole and electron doping in different parts of the channel. Figure 3a and Figure 3b show false-color maps of the magnitude of $I_{ds}$ at a fixed bias voltage as the split gates are independently swept from 7 to 12 V. The gate voltage values are chosen to be close to the flatband condition as determined from the transfer characteristics measured in single back-gate configuration (Figure 2a). In both maps, it is possible to distinguish four distinct regions where current flows, separated by regions where only a small current is measured (the OFF state of the device). The previously reported effect of gate-induced stress for devices on $SiO_2$ is drastically reduced by choosing a gate dielectric that provides a more stable dielectric environment (h-BN), as can be seen from the small (~ 4 V) shift in the flatband voltage with respect to Figure 2a. Moreover, the shift in the gate voltage is constant over several days, in contrast with the much shorter time scales (minutes) of devices fabricated on $SiO_2$.[26]

The cross of supressed conductance allows us to identify four distinct regions of doping, indicating the good control of the gates over the doping type and concentration in the b-P flake. In Figure 3a, the NN ($V_{lg} = V_{rg} = 12$ V) and PP ($V_{lg} = V_{rg} = 7$ V) quadrants show the highest current values, as expected from the transfer characteristics in Figure 2a. On the other hand, the measured current in the NP ($V_{rg} = 7$ V, $V_{lg} = 12$ V) region is much higher than in the PN ($V_{rg} = 12$ V, $V_{lg} = 7$ V) region, suggesting the formation of a junction diode when the gates are biased with opposite polarities. To confirm that the asymmetry in the current is due to the formation of a PN or NP junction, we repeat the measurement at negative bias (Figure 3b). Again, the NN and PP region show the highest current. However, the current in the NP configuration is now much smaller than the current in the PN configuration, confirming the formation of a junction diode as different type of charge carriers are accumulated in the channel.

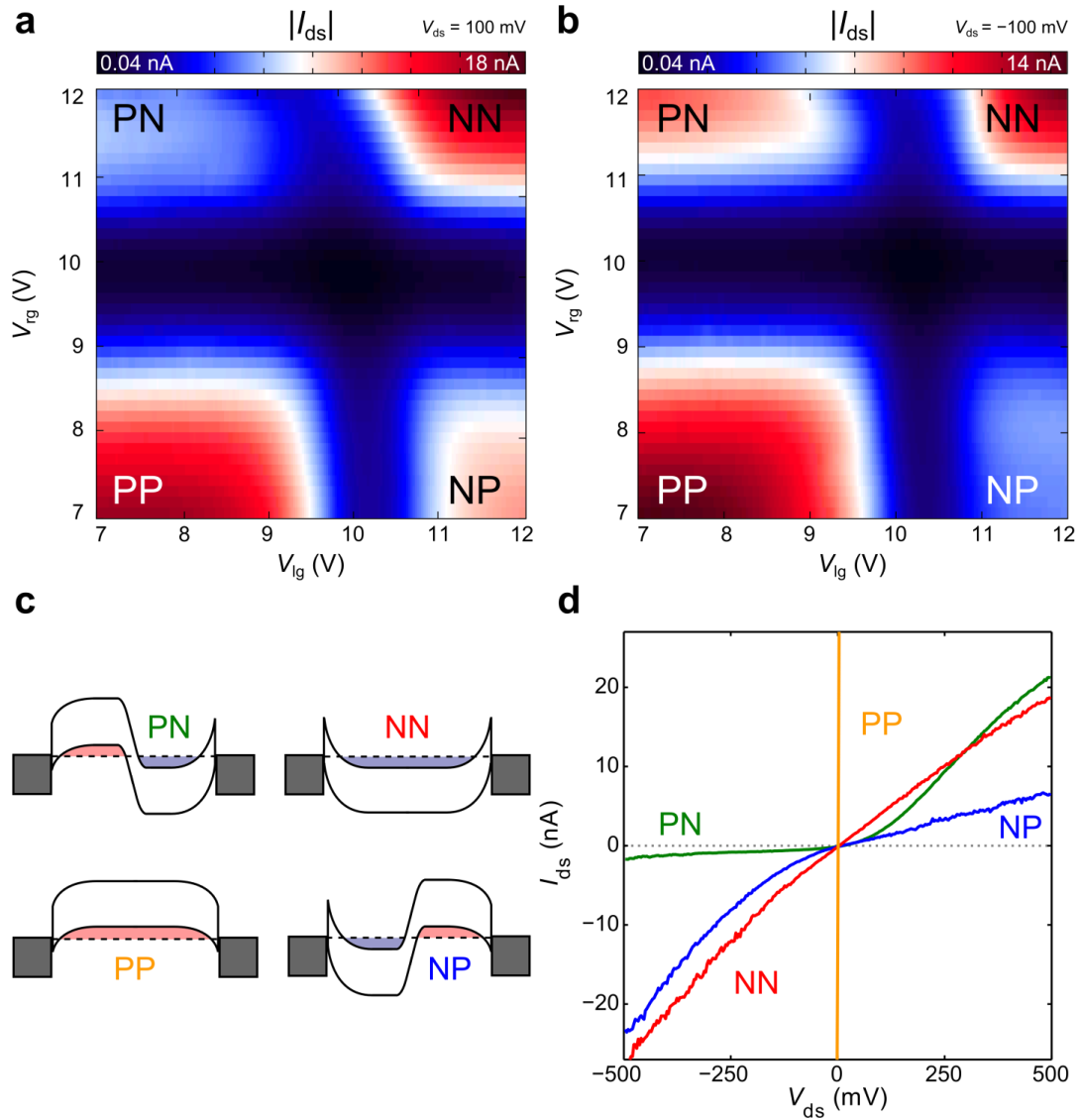

**Figure 3** (**a**) False color map of the magnitude of the source-drain current at constant bias voltage (+ 100 mV) as the voltage on the two local gates is changed independently. (**b**) False color map of the magnitude of the source-drain current at constant bias voltage (- 100 mV) as the voltage on the two local gates is changed independently. (**c**) Band diagrams corresponding to the different device configurations (**d**) Output characteristics of the device in different gate configurations (PP: $V_{lg} = V_{rg} = -10$ V, NN: $V_{lg} = V_{rg} = 10$ V, NP: $V_{lg} = 10$ V, $V_{rg} = -10$ V, PN: $V_{lg} = -10$ V, $V_{rg} = 10$ V).

The origin of the strong current modulation induced by the local gates can be explained by the band diagrams in Figure 3c. In the NN and PP regions, the electric field generated by the two gates is homogeneous and favours either electron (NN) or hole (PP) conduction. As the gates are biased in opposite polarities (PN or NP configuration), a strong horizontal electric field is also present in the b-P region bridging the gates. This large horizontal electric field contributes to charge accumulation in the two different regions of the b-P flake, giving rise to an asymmetric energy landscape. Therefore, the horizontal electric field generated by the local gates and the charge accumulation in the different regions of the b-P flake generate a PN or NP junction, which is then responsible for the measured rectifying output characteristics (see Figure S9 in the Supplementary Material). This is similar to a PN junction in bulk-doped materials in which the dopant charges are now located at the surface of the gates rather than in the bulk of the semiconductor itself.

The control of the electrical response of the device with the local gates is even more evident by looking at the measured output characteristics in the different doping configurations (Figure 3d). In the NN and PP configuration, the $I_{ds}$-$V_{ds}$ curves are linear, as expected from Figure 2b and 2c. In the PN and NP configuration, the $I_{ds}$-$V_{ds}$ curves are highly non-linear and the device shows rectifying behaviour, whose direction can be controlled by the gate bias polarity. In the PN configuration, the reverse-bias current is in the order of 1 nA ($V_{ds}$ = −500 mV), while the reverse-bias current in the NP configuration is around 5 nA ($V_{ds}$ = 500 mV). We can attribute the non-zero reverse-bias current in both PN and NP configuration to leakage current due to thermally excited carriers. Given the small bandgap of few-layer b-P, the leakage current is higher than in PN junctions based on large-bandgap 2D semiconductors.[30,31].

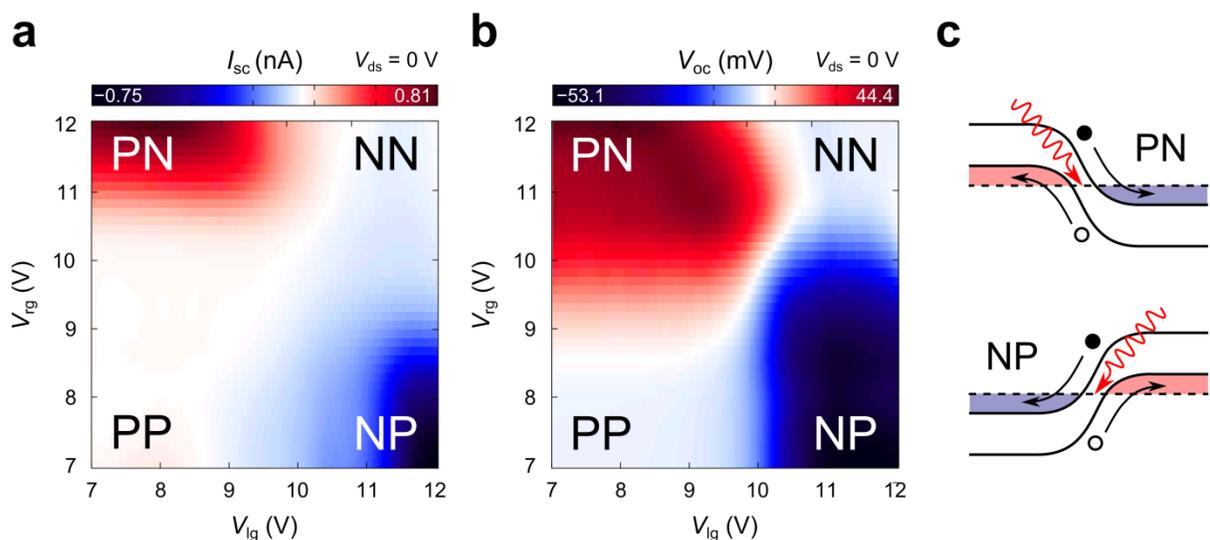

**Figure 4** (**a**) False color map of the short circuit current ($I_{sc}$, $V_{ds} = 0$ V) as the voltages on the two local gates are changed independently. (**b**) False color map of the open circuit voltage ($V_{oc}$, $I_{ds} = 0$ A) as the voltages on the two local gates are changed independently. (**c**) Band diagrams illustrating the photovoltaic mechanism in PN and NP configuration.

A useful feature of PN junctions is the large built-in electric field that can be used to separate photoexcited carriers *via* the photovoltaic effect.[32-34] Upon illumination, the internal electric field at the junction separates photogenerated carriers and gives rise to a photocurrent at zero external bias (short-circuit current, $I_{sc}$) and a photovoltage with no current flowing (open circuit voltage, $V_{oc}$). Figure 4a shows a false-color map of $I_{sc}$ of the device under illumination ($\lambda = 640$ nm, $P = 3.9$ μW) as the gates voltages are swept independently. Four regions are clearly visible: high photocurrent in PN and NP configuration and no photocurrent in the PP and NN configuration. Figure 4b displays the measured photovoltage ($V_{oc}$); again four distinct regions appear: high $V_{oc}$ in PN and NP configuration and zero $V_{oc}$ in PP and NN configuration. The photocurrent and photovoltage are only generated when the local gates are oppositely biased, indicating that the mechanism is photovoltaic originating from the formation of a PN or NP junction in the channel.

Figure 4c shows schematically that when the gates are biased in opposite polarities, the internal electric field in the b-P channel separates photogenerated carriers under above-bandgap illumination, generating a photocurrent. If the circuit is kept open, the photogenerated carriers accumulate in distinct parts of the flake, giving rise to a photovoltage. This demonstrates that the b-P PN junction can be used for energy harvesting.

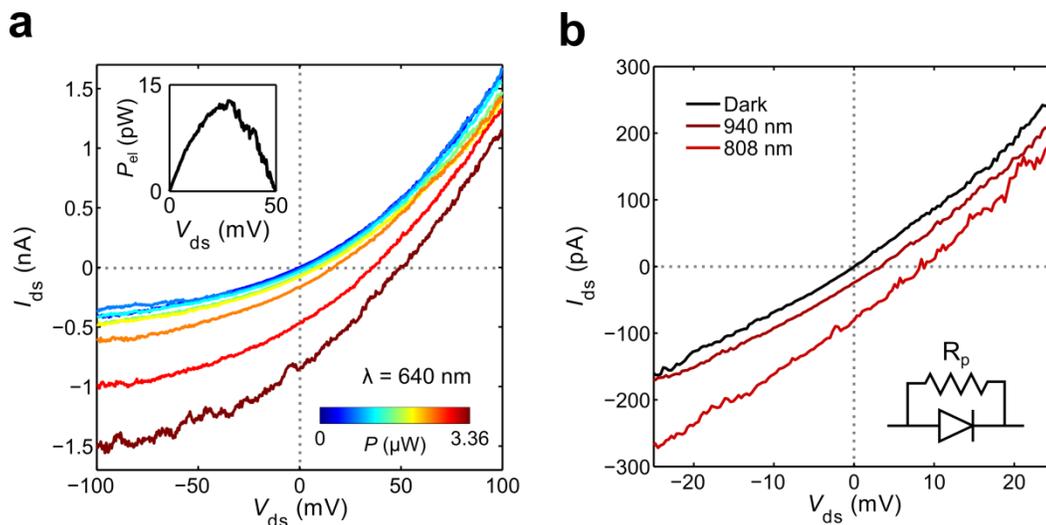

**Figure 5** (**a**) Output characteristics in the PN configuration as a function of the incident optical power ($\lambda = 640$ nm). The inset shows the electrical power that can be harvested at the maximum employed illumination power. (**b**) Output characteristics in the PN configuration in dark (black solid line) and under illumination of different wavelengths excitation wavelength at fixed power ($P = 0.33$ µW). The inset shows the schematics of the equivalent circuit.

In figure 5a, we characterize the optoelectronic properties of few-layer b-P junction diodes by measuring the output characteristics under illumination with varying incident power (Figure 5a). As the power is increased, the $I_{ds}$-$V_{ds}$ characteristics shift towards more negative current values. This is consistent with a photocurrent generation mechanism where the separation of the electron-hole pairs is driven by the internal electric field at the junction region: the photocurrent has the same sign as the reverse-bias current; therefore, under illumination and reverse-bias, the total measured current is larger than in dark. Thus, an additional forward-bias is required to compensate the photogenerated reverse-current, giving rise to a non-zero open-circuit voltage. The measured $I_{sc}$ is in the order of 1 nA and $V_{oc}$ ~ 50 mV at the highest incident power and both $I_{sc}$ and $V_{oc}$ are linear as function of the incident power (Figure S7). Moreover, we can evaluate the electrical power harvested by the b-P PN junction by $P_{el} = I_{ds} \cdot V_{ds}$ (Figure 5a, inset) and find that the maximum $P_{el}$ is in the order of 13 pW, in good comparison with other electrostatically defined PN junctions based on $WSe_2$.[33,34]

While the photocurrent is in the same order of magnitude as similar devices realized with single layer $WSe_2$,[33,34] the open circuit voltage is lower, due to the smaller bandgap of few-layer b-P. We note that the measured $V_{oc}$ is significantly lower than the predicted bandgap of few-layer b-P (~ 300 meV). We attribute this to a resistance in parallel ($R_p$) to the junction (see inset of Figure 5b) which limits $V_{oc}$ by leakage of photogenerated carriers across the junction. The slope of the $I_{ds}$-$V_{ds}$ curve at $V_b = 0$ V (in dark, Figure 5b) is a signature of this parallel resistance.

The b-P PN junctions show photovoltaic effect up to the near-infrared (NIR) part of the electromagnetic spectrum. Figure 5b plots the $I_{ds}$-$V_{ds}$ curves in PN configuration in dark (solid black line) and with excitation wavelengths of 808 nm, 885 nm and 940 nm ($P = 0.33$ μW). Under illumination, the output characteristics confirm that the photocurrent generation mechanism is dominated by the photovoltaic effect. As the photon energy is increased, both $I_{sc}$ and $V_{oc}$ increase, consistent with increased absorption in the b-P flake at shorter wavelengths. The measured photovoltaic effect up to 940 nm illumination indicates that the bandgap of the b-P flake is smaller than 1.31 eV and demonstrates energy harvesting in NIR part of the spectrum which constitutes a strong advantage with respect to other 2D semiconductors such as Mo- or W-chalcogenides whose large bandgap (> 1.6 eV) limits their applicability in the NIR.[13,35,36]

In conclusion, we have fabricated PN junctions based on van der Waals heterostructures of two different 2D materials: h-BN as gate dielectric and b-P as ambipolar semiconducting channel material. We demonstrate full electrostatic control of the device by means of local gating which allows us to tune the electrical behaviour of the device from metallic to rectifying. We observe a strong photocurrent and a significant open-circuit photovoltage, which we attribute to electron-hole separation at the PN junction from the photovoltaic effect, which extends even up to the NIR. Our work sets the ground for further research towards broad-band energy harvesting devices based on few-layer black-phosphorus.

*Methods*

*Fabrication of black Phosphorus (b-P) PN junctions.*

Vertical heterostructures of hexagonal boron nitride (h-BN) and b-P were fabricated by means of an all-dry transfer technique, as described in Refs. [18,37]. We start from a pre-patterned $SiO_2$ (285nm) /Si substrate with a pair of split gates (Ti/AuPd 5 nm/ 60 nm) fabricated by e-beam lithography. We then exfoliate h-BN from a powder (Momentive, Polartherm grade PT110) onto a viscoelastic stamp (GelFilm® by GelPak) with blue Nitto tape (Nitto Denko Co., SPV 224P). Thin h-BN flakes are identified by optical inspection and stamped onto the split gate pair by gently pressing the viscoelastic stamp surface against the substrate and then slowly releasing it. Afterwards, few layer b-P flakes are identified by the same method and transferred on top of the h-BN flake. The leads (Ti – 2 nm/Au – 60 nm) are then patterned with e-beam lithography (*Vistec, EBPG5000PLUS HR 100*), metal deposition (*AJA international*) and lift-off (warm acetone).

Atomic Force Microscopy (AFM) is used to determine the thickness of h-BN and b-P flakes. The AFM (*Digital Instruments D3100 AFM*) is operated in amplitude modulation mode with Silicon cantilevers (spring constant 40 N m$^{-1}$ and tip curvature <10 nm) to measure the topography and to determine the number of b-P layers.

*Opto-electronic characterization*

All the opto-electronic characterization is performed in a *Lakeshore Cryogenics* probestation at room temperature under vacuum (~ 10$^{-5}$ mBar). The light excitation is provided by diode pumped solid state lasers operated in continuous wave mode (CNI Lasers). The light is coupled into a multimode optical fiber (NA = 0.23) through a parabolic mirror ($f_{\text{reflected}}$ = 25.4 mm). At the end of the optical fiber, an identical parabolic mirror collimates the light exiting the fiber. The beam is then directed into the cryostat. The beam spot size on the sample has a diameter of 230 ± 8 μm for all wavelengths. The power quoted in the main text is the power incident on the device area (*P*) calculated as follows $P = P_{laser} \frac{A_{device}}{A_{spot}}$, where $P_{\text{laser}}$ is the power of the laser, $A_{\text{device}}$ is the area of the device and $A_{\text{spot}}$ is the area of the laser spot.

*Author information*

Michele Buscema and Dirk J. Groenendijk contributed equally to this work.

*Affiliations*

Kavli Institute of Nanoscience, Delft University of Technology, Lorentzweg 1, 2628 CJ Delft (The Netherlands).

*Contributions*

D.J.G fabricated the devices and conducted the measurements. M.B. and D.J.G. co-wrote the manuscript. M.B. and A.C-G. designed the experiments. A.C-G directed the research project. All authors discussed the data and gave feedback during the elaboration of the manuscript.

*Acknowledgements*

This work was supported by the European Union (FP7) through the program RODIN and the Dutch organization for Fundamental Research on Matter (FOM). A.C-G. acknowledges financial support through the FP7-Marie Curie Project PIEF-GA-2011-300802 ('STRENGTHNANO').

**Supporting Information**

**Photovoltaic effect in few-layer black phosphorus PN junctions defined by local electrostatic gating**

Michele Buscema,[†] Dirk J. Groenendijk[†], Gary A. Steele, Herre S.J. van der Zant and Andres Castellanos-Gomez*

Kavli Institute of Nanoscience, Delft University of Technology, Lorentzweg 1, 2628 CJ Delft (The Netherlands).

Table of contents:

**Electrical characterization of two other fabricated devices**

Two other locally gated b-P flakes were measured. In Figure S1a an optical micrograph of a second device is shown. Figure S1b displays the measured transfer characteristics ($I_{ds}$-$V_g$ at fixed $V_{ds}$). This device also displays strong ambipolar behaviour, with $I_{ds}$ varying from a few µA (hole-enhancement) to ~ 500 nA (electron enhancement) when sweeping the linked gates from -15 to 15 V. The inset shows the data in a linear scale. We extract a field effect mobility of 17 cm$^2$/Vs for holes. Figure S1c shows the output characteristics of the device in different gate configurations. The IVs are linear in PP ($V_{lg}$ = $V_{rg}$ = -15 V) and NN ($V_{lg}$ = $V_{rg}$ = 15 V) configuration. However, the IVs show strongly rectifying behaviour when the gates are biased at different polarities (PN and NP configuration).

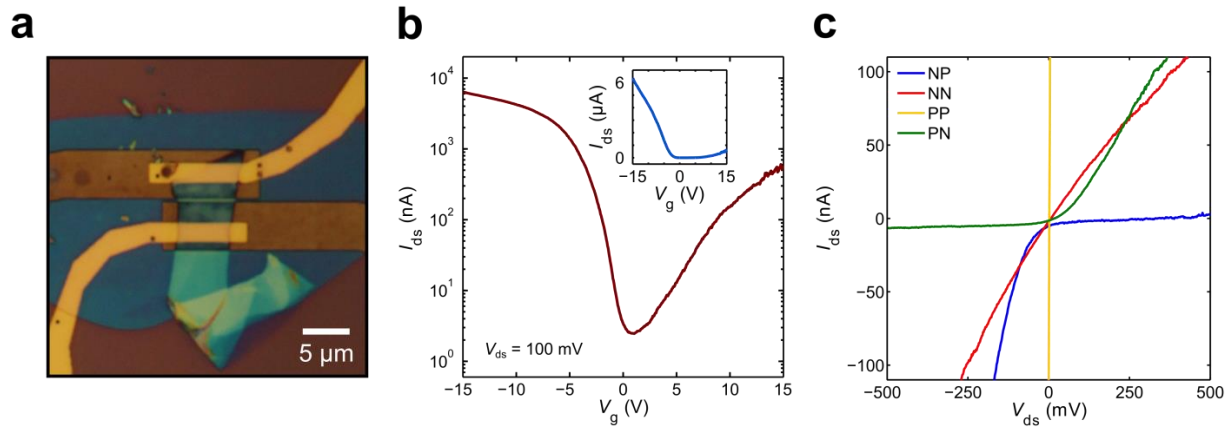

**Figure S1.** (a) Optical micrograph of the device. (b) Transfer characteristic at $V_{ds}$ = 100 mV. The inset shows the data in a linear scale. (c) Output characteristics of the device in different gate configurations (PP: $V_{lg}$ = $V_{rg}$ = -15 V, NN: $V_{lg}$ = $V_{rg}$ = 15 V, NP: $V_{lg}$ = 15 V, $V_{rg}$ = -15 V, PN: $V_{lg}$ = -15 V, $V_{rg}$ = 15 V).

Figure S2a shows an optical micrograph of a third locally gated PN junction. Note that the diagonal line across the channel is a part of the flake which is folded or elevated from the boron nitride. The transfer characteristics are shown in Figure S2b. The inset shows the data in a linear scale. We extract a field effect mobility of 0.27 cm$^2$/Vs for holes. Figure S2c shows the output characteristics of the device in different gate configurations. The PP and NN configuration show linear behaviour, whereas the PN configuration shows diode-like behaviour. The IV in the NP configuration does not show rectifying behavior; this might be related to imperfections of the b-P flake as can be seen in Figure S2a.

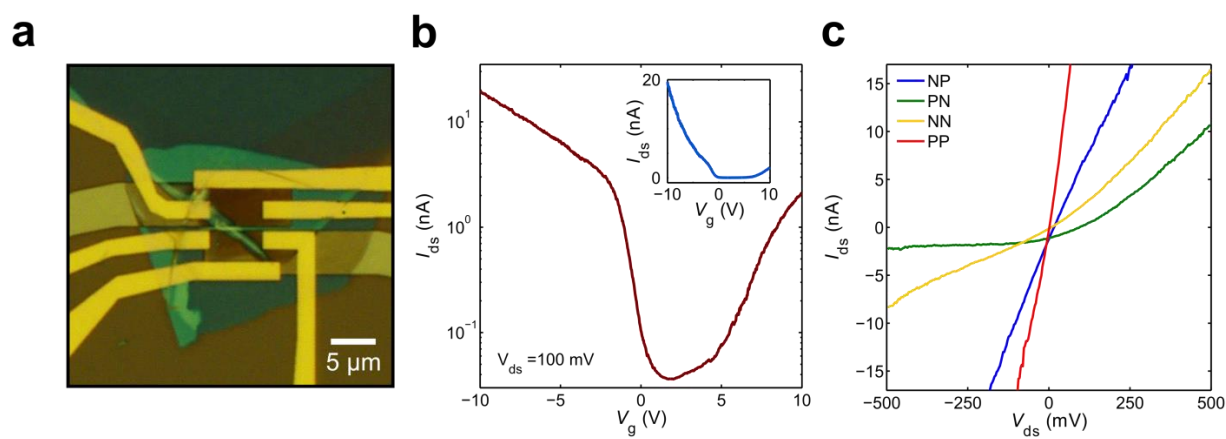

**Figure S2** (a) Optical micrograph of the device. (b) Transfer characteristic at $V_{ds}$ = 100 mV. The inset shows the data in a linear scale. (c) Output characteristics of the device in different gate configurations (PP: $V_{lg} = V_{rg} = -10$ V, NN: $V_{lg} = V_{rg} = 10$ V, NP: $V_{lg} = 10$ V, $V_{rg} = -10$ V, PN: $V_{lg} = -10$ V, $V_{rg} = 10$ V).

**Topography of the studied devices**

Figure S3a-c show AFM (atomic force microscopy) images of the three measured devices. Figure S3d-f show the corresponding optical images.

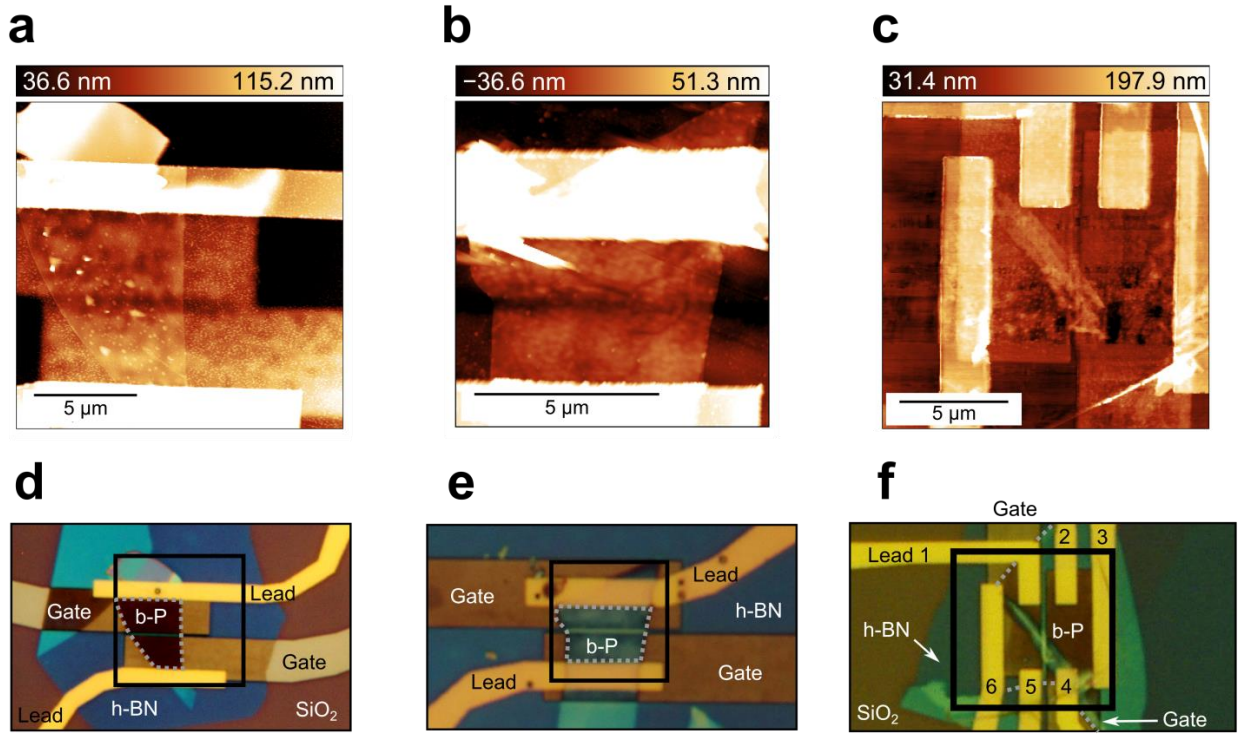

**Figure S3.** (a) AFM of the device described in the main text. b-P thickness ~ 6 nm, h-BN thickness ~ 20 nm. (b) AFM of the device presented in Figure S1. b-P thickness ~ 14.5 nm, h-BN thickness ~ 15 nm. (c) AFM of the device presented in Figure S2. b-P thickness ~ 5 nm, h-BN thickness ~ 25.8 nm. (d) Optical image of the device described in the main text. (e) Optical image of the device presented in Figure S1. (f) Optical image of the device presented in Figure S2. In panels (d) to (f) the black box indicate the region where the AFM image is measured.

**Electrical characterization of two other fabricated devices under illumination**

Figure S4a shows the $I_{ds}$-$V_{ds}$ characteristics in PN configuration ($V_{lg}$ = -4 V, $V_{rg}$ = 9 V) of the device presented in Figure S1 in dark and under illumination ($\lambda$ = 640 nm, $P$ = 3.03 µW). The inset shows the data in a smaller bias range (from -20 mV to 20 mV). The shape of the IV under illumination may be due to a combination of processes: photovoltaic (causing an open-circuit voltage and short-circuit current) and photoconductive (changing the slope in reverse and forward bias). The photoconductive component might be caused by conduction in the top region of the flake, which is not affected by the gates due to screening (since the flake is thicker). Figure S4b shows the $I_{ds}$-$V_{ds}$ characteristics in PN configuration ($V_{lg}$ = -10 V, $V_{rg}$ = 10 V) of the device presented in Figure S2 in dark and under illumination ($\lambda$ = 640 nm, $P$ = 2.60 µW). The inset shows the data in a smaller bias range (from -150 mV to 150 mV). The IV under illumination preserves the diode-like character, indicating that the photovoltaic process is dominant.

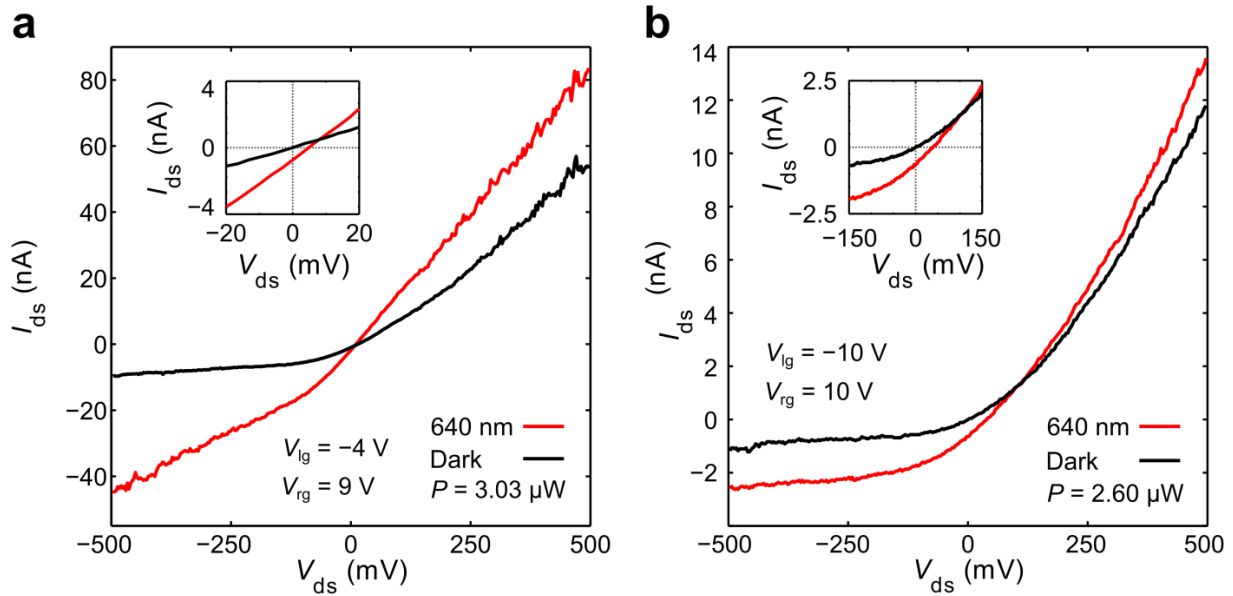

**Figure S4** (a) Device presented in Figure S1. The gates are biased in different polarities ($V_{lg}$ = -4 V, $V_{rg}$ = 9 V). The inset shows the data in a smaller bias range. (b) Device presented in Figure

S2. . The gates are biased in different polarities ($V_{lg}$ = -10 V, $V_{rg}$ = 10 V). The inset shows the data in a smaller bias range.

**Electrical characterization of the device in Figure S1 under illumination with different wavelengths**

Figure S5a shows $I_{ds}$-$V_{ds}$ characteristics (in PN configuration) for the device presented in Figure S1 under illumination by different wavelengths. Qualitatively, the presented IVs show similar behavior, showing an increase in current in forward and reverse bias. The responsivity $R$ can be calculated by $R = I_{ph}/P$, where $I_{ph}$ is the photocurrent and $P$ is the illumination power. The responsivity $R$ at $V_{ds}$ = -500 mV is plotted as a function of wavelength in Figure S5b. The responsivity increases up to 28 mA/W as the wavelength decreases. This value is higher than previously reported for devices on $SiO_2$.[1]

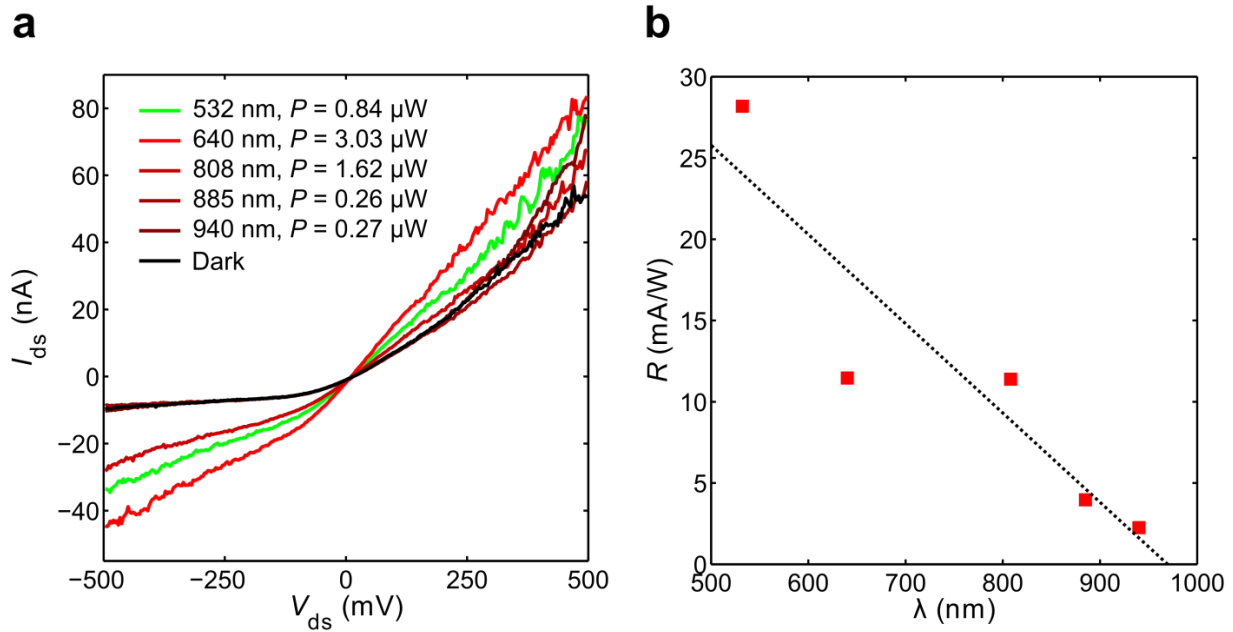

**Figure S5.** (a) Output characteristics for wavelengths ranging from 532 to 940 nm. (b) Responsivity as a function of wavelength extracted from Figure S6a at $V_{ds}$ = -500 mV.

**Electrical characterization of the device in Figure S2 under illumination with different power**

Figure S6a shows the Ids-Vds characteristics (in PN configuration) near zero bias for different illumination powers ($\lambda = 640$ nm) for the device presented in Figure S2. The generated electrical power P can be calculated by $P_{el} = V \cdot I$. The magnitude of the generated electrical power is plotted in Figure S6b. The generated electrical power reaches a maximum of ~ 2.7 pW for illumination power 2.82 µW. Figure S6c shows the open circuit voltage and the short circuit current against the illumination power, confirming the linear relationship with power found for the device presented in the main text.

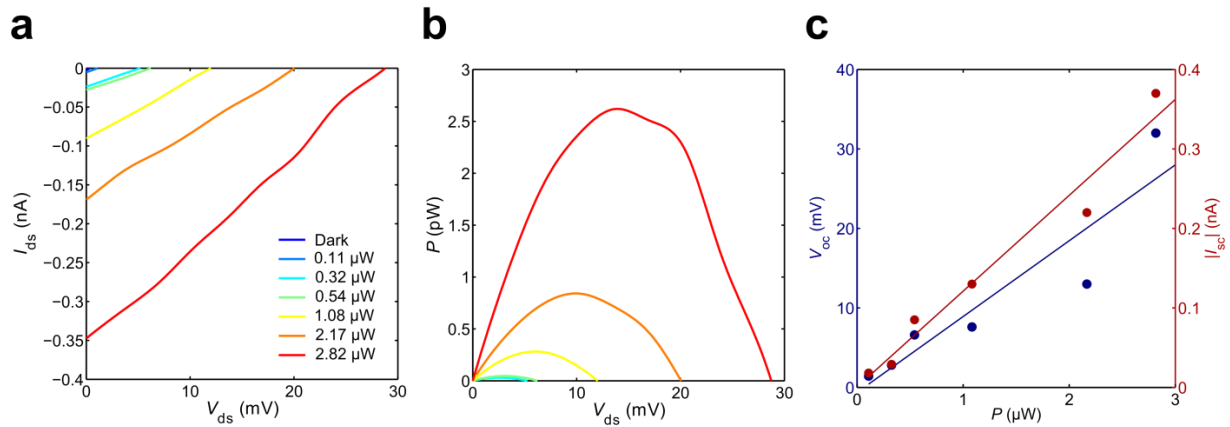

**Figure S6.** (a) Ids-Vds characteristics for different illumination powers near zero bias. (b) Generated electrical power for different illumination powers. (c) $V_{oc}$ (left axis, blue dots) and $I_{sc}$ (right axis, red dots) against illumination power.

**Electrical characterization of the device in the main text under illumination with different power**

Figure S7 displays $V_{oc}$ (left vertical axis) and $I_{sc}$ (right vertical axis) against the optical power incident on the device. $V_{oc}$ increases linearly with the incident optical power. In an ideal case, the open-circuit voltage should increase logarithmically with the optical power incident on the device. The deviation from the ideal behaviour can be attributed to parasitic resistive losses in our device. $I_{sc}$ is linear with the excitation power for the whole range of incident optical power, indicating that the recombination rate of photogenerated carrier is independent from the rate of incoming photons. This is usually associated with monomolecular recombination: the recombination rate is linearly proportional to the density of photogenerated carriers, even in the presence of trap states.[2,3]

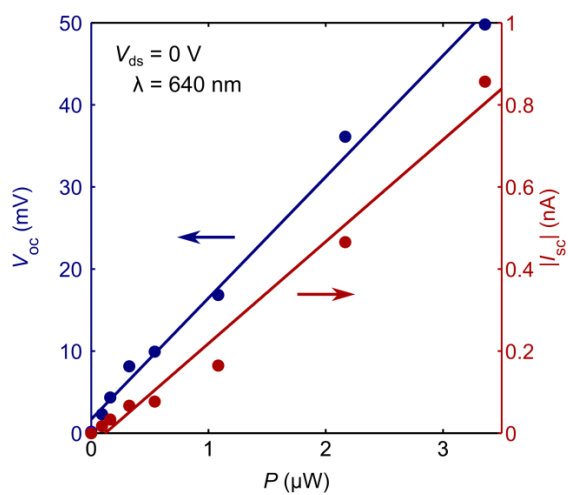

**Figure S7.** Open circuit voltage (left axis) and short-circuit current (right axis) as function of incident optical power.

**Electrical characterization of the device in the main text under illumination with different wavelengths**

Figure S8a shows $I_{ds}$-$V_{ds}$ characteristics (in PN configuration) for the device presented in the main text under illumination by different wavelengths. Figure S8b shows the same data close to zero bias. Similar to Figure S5, the generated electrical power is calculated and plotted in Figure S7c, reaching a maximum of around 0.27 pW for illumination with P = 0.32 µW and $\lambda$ = 640 nm. Note that the incident optical power used in this measurement is more than 10 times lower than the power used in Figure 5 in the main text.

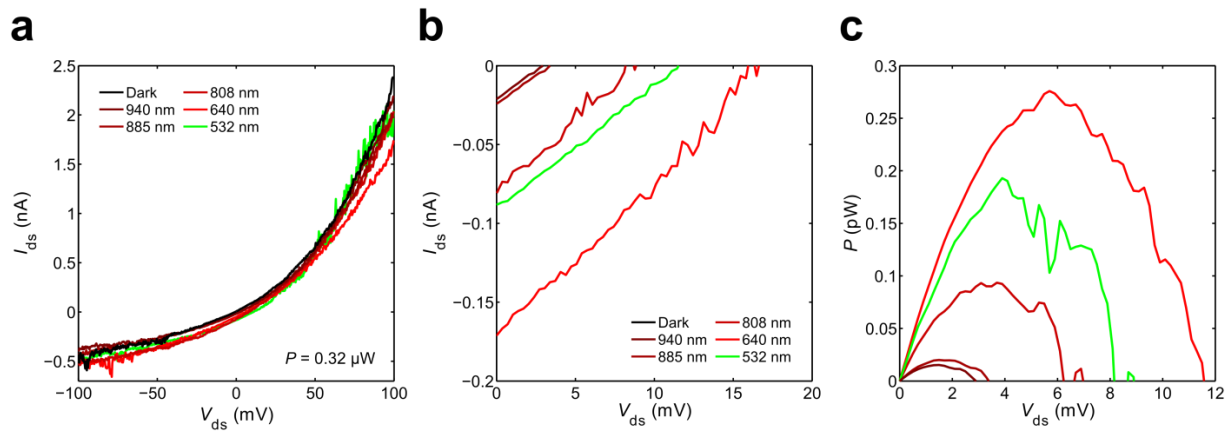

**Figure S8** (a) Output characteristics for wavelengths ranging from 532 to 940 nm. (b) Output characteristics near zero bias, showing photovoltaic behavior. (c) The magnitude of the generated power as a function of $V_{ds}$, obtained by multiplying the voltage values in Figure S7b by the current values.

**Modelling the formation of an electrostatically defined PN junction.**

The formation of a PN junction in a semiconductor by local gating was studied by simulations performed with COMSOL Multiphysics. Figure S9a is a schematic (in scale) of the geometry used for the simulation. The different components are indicated: the local gates ($\varepsilon_r$ gold, yellow) with hexagonal boron nitride ($\varepsilon_r = 4$, blue) and black phosphorus ($\varepsilon_r$ b-P, gray), on top. The surrounding regions are vacuum ($\varepsilon_r = 1$). The left gate is set to +10 V and the right gate is set to − 10 V. Figure S9b shows the resulting electric field lines. Figure S9c shows the calculated electric potential at the bottom of the b-P flake. Above the left gate the potential is + 10 V, smoothly decreases across the junction region and reaches -10 V above the right gate. The magnitude of the horizontal electric field $E_x$ is given by $|\frac{dV}{dx}|$ and is shown in Figure S9d. In the junction region, $E_x$ is positive, confirming that the electric field is oriented towards the right. The horizontal electric field will make electrons drift to the left and holes to the right, contributing to charge accumulation above the gates and will cause depletion of charges in the junction region.

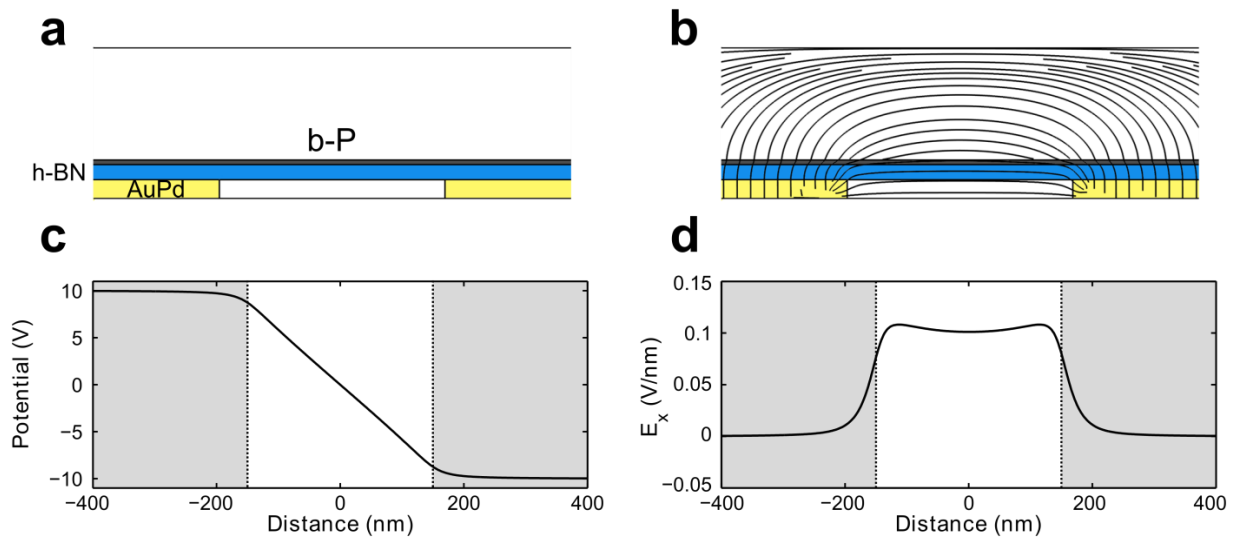

**Figure S9 (a)** Simulation geometry indicating the different materials. The gap between the gates is 300 nm wide and the plot is centered around zero. (b) Calculated electric field lines across the device. (c) Linecut of the electric potential along the junction region. (d) Magnitude of the horizontal electric field $E_x$ along the junction. The electric field is positive in the junction region indicating that it is oriented to the right.

Supplementary References